\documentclass{aastex}          
\usepackage{spr-astr-addons}    
\usepackage{url}

\begin{document}
%
\title{Recent advances in modeling stellar interiors}

\shorttitle{Modeling stellar interiors}
\shortauthors{J. A. Guzik}

\author{Joyce Ann Guzik}{\altaffilmark{1}} 

\altaffiltext{1}{Los Alamos National Laboratory, Los Alamos, NM 87545 USA  joy@lanl.gov}


\begin{abstract}

Advances in stellar interior modeling are being driven by new data from large-scale surveys and high-precision photometric and spectroscopic observations.  Here we focus on single stars in normal evolutionary phases; we will not discuss the many advances in modeling star formation, interacting binaries, supernovae, or neutron stars.  We review briefly:  1) updates to input physics of stellar models; 2) progress in two and three-dimensional evolution and hydrodynamic models; 3) insights from oscillation data used to infer stellar interior structure and validate model predictions (asteroseismology).  We close by highlighting a few outstanding problems, e.g., the driving mechanisms for hybrid $\gamma$ Dor/$\delta$ Sct star pulsations, the cause of giant eruptions seen in luminous blue variables such as $\eta$ Car and P Cyg, and the solar abundance problem.

\end{abstract}


\section{Introduction}

Advances in stellar modeling are being driven and validated by stellar oscillation data.  Over the last twenty years, multiple oscillation modes of many star types in all evolutionary phases have been discovered and observed from ground-based networks and satellites.  These modes include gravity and pressure modes, radial and nonradial modes (see, e.g., Christensen-Dalsgaard \& Houdek 2009). 

Here we limit the discussion to single stars (neglecting binary interactions), stellar interiors only (neglecting atmosphere models and winds), and normal phases of evolution (neglecting star formation, supernovae, brown dwarfs, neutron stars, and black holes).

\section{Physics of stellar interior models}

\subsection{Stellar structure equations}

The equations for one-dimensional stellar structure and evolution modeling, neglecting rotation and magnetic fields, comprise four equations for conservation of mass, hydrostatic equilibrium, conservation of energy, and energy transport (see., e.g, Clayton 1983).  In Lagrangian (mass) coordinates, these are:\\


$\frac{dr}{dM} = \frac{1}{4 \pi r^2 \rho} $\\

$\frac{dP}{dM} = -{\frac{GM_r}{4 \pi r^4}} $\\

$\frac{dL}{dM} = \varepsilon - T\frac{dS}{dt}$\\

$\frac{dT}{dM} =-{\frac{3}{4ac}} \frac{\kappa}{T^3} \frac{L_r}{16\pi^2 r^4}$\\

or, if $\nabla_{rad} > \nabla_{ad}$,\\

$\frac{dT}{dM} =\frac{\Gamma_2 -1}{\Gamma_2} \frac{T}{P} \frac{dP}{dM}$\\

where\\

 $\frac{\Gamma_2 - 1}{\Gamma_2} = ( \frac{\partial ln T}{\partial ln P} )_S \equiv  \nabla _{ad}$\\


The symbols refer to mass $M$, temperature $T$, pressure $P$, entropy $S$, luminosity $L$, radiative opacity $\kappa$, time $t$, radiation constant $a$, speed of light $c$, nuclear energy generation rate $\varepsilon$, and subscripts $rad$ and $ad$ refer to the radiative and adiabatic gradients, respectively.  To solve these equations in a stellar evolution code also requires input for the assumed element abundances and corresponding opacities and equation of state, nuclear reaction rates, a treatment for convection, and a surface boundary condition assumption or a model atmosphere.  In some cases it is also critical to account for diffusive settling or radiative levitation of elements, mass loss in massive stars and late evolution stages, rotationally-induced mixing, and convective overshooting.  In addition, magnetic fields and MHD effects, turbulence driven by instabilities, turbulent pressure and energy, mixing and momentum and energy transport due to acoustic or gravity waves, pulsations and pulsation-convection interactions, and stellar winds are not modeled explicitly.  Many of these processes require multi-dimensional modeling, or produce effects on timescales significantly shorter than the evolution timescale. Sometimes, a parameterized treatment of these processes is included in 1D evolution models, often based on 2D and 3D dynamical models.  

\subsection {Opacities}

Radiative opacities commonly in use today are included in the form of tables calculated for particular abundance mixtures and interpolated.  These are available from the following groups:  Lawrence Livermore National Laboratory OPAL (Iglesias \& Rogers 1996) ; OP (Seaton 2005; Badnell et al. 2005);  OPAS from CEA, France (see Blancard et al., these proceedings); and Los Alamos National Laboratory LEDCOP (Magee et al. 1995).  The Rosseland mean opacities for stellar interior conditions usually agree among these groups to better than 10\%, and often to within a few percent,  but for many stellar astrophyiscs problems even this small difference can affect evolution and interpretation of stellar oscillation data.  For the sun and some other types of pulsating stars, opacity increases would improve agreement with data.


For temperatures $<$ 6000 K near the stellar photosphere, these opacities must be supplemented by low-temperature tables including effects of molecular lines and dust formation.  The Ferguson et al. (2005) update to the Alexander \& Ferguson (1994) tables are now most  in use;  their update includes 31 vs. 6 species of dust, 800 million vs. 30 million lines, 24,000 vs. 9000 wavelengths, a more complex equation of state, and updated optical constants.  It spans temperatures from 500 to 30,000 K, and agrees well with OPAL in the overlap region.  As seen in Fig. 1 (see Guzik et al. 2005 and Guzik \& Mussack 2010), the increase in opacity for the updated low-temperature opacity tables is significant for the higher frequency solar oscillations that are sensitive to conditions near the photosphere.



 \begin{figure}
 \includegraphics[width=\columnwidth]{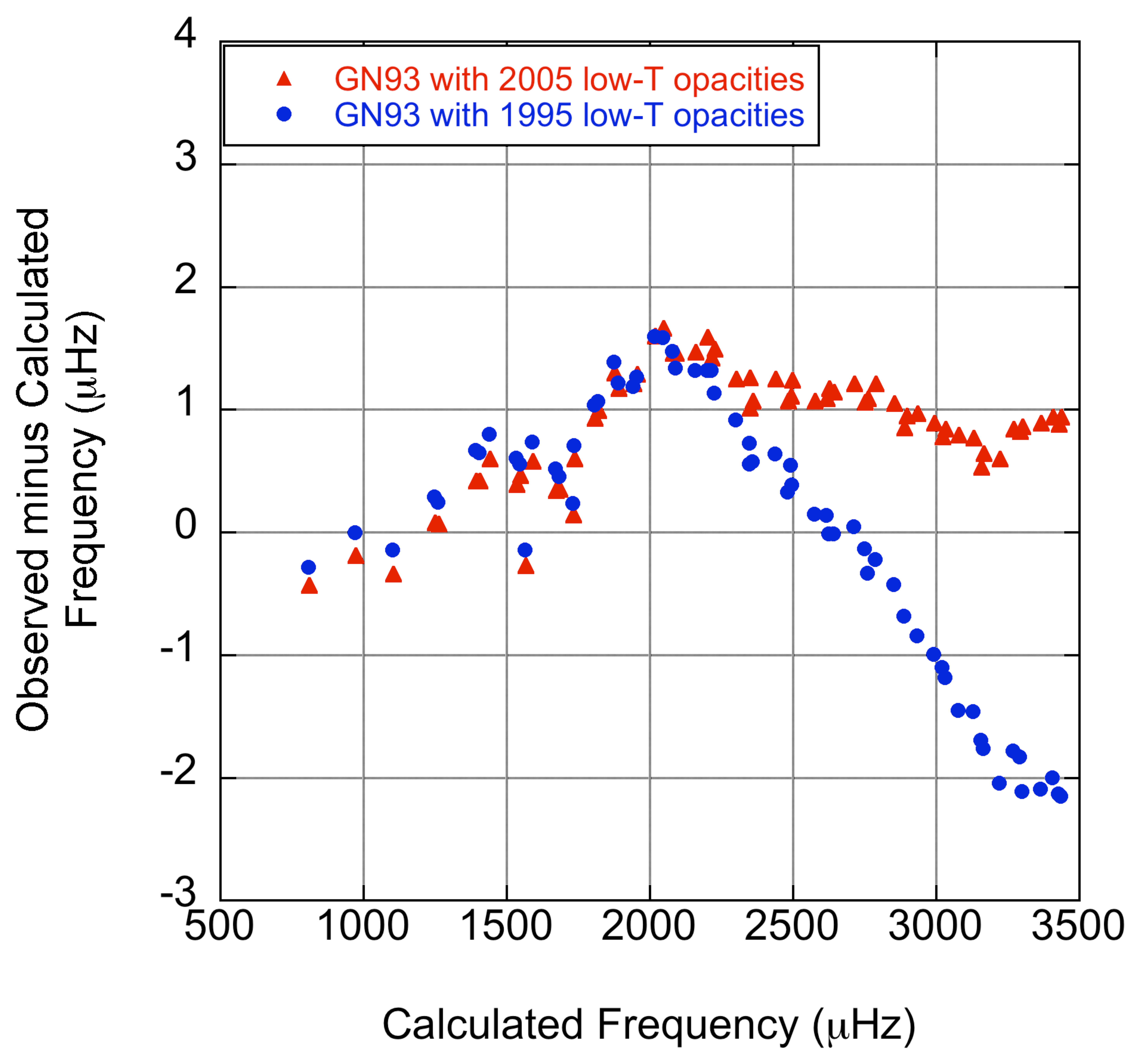}
 \caption{Observed minus calculated (O-C) nonadiablatic frequencies for degrees $\ell$ = 0, 2, 10, and 20 of solar models with Grevesse \& Noels (1993) element abundances using 2005 and 1995 low-temperature opacities.  See Guzik et al. (2005) and Guzik \& Mussack (2010) for details.  The trend in O-C frequency is flatter for the models with the  2005 opacities.} 
\label{fig:1}
\end{figure}

Electron thermal condition can be a significant souce of energy transport, particularly for the denser part of stellar interiors, white dwarfs, and low-mass stars.  Conduction can be taken into account by adding the conductive opacity in reciprocal to the Rosseland mean opacity.  The contribution is 1-2\% for the  solar interior.  For a long time, the Hubbard \& Lampe (1969) formula was used to calculated conductive opacities.  These were replaced by the Itoh et al. (1984, 1993a, b; see also 1994 errata) formulas.  Later Potekhin et al. (1999) proposed a significant revision.  However, Cassisi et al. (2007) revise the Potekhin et al. work to include electron-electron scattering and partial degeneracy.  They show that the new results for white dwarfs and low-mass stars are very close to those obtained using the Hubbard \& Lampe (1969) formulation.

\subsection {Equation of state}

An equation of state model is needed to obtain internal energy and pressure given temperature and density, as well as calculate various thermodynamic quantities and derivatives for other physical models, e.g. the convection model.  Both analytical treatments and tables have been used.  Analytical treatments may be more computationally intensive, but allow for smoother derivatives that may be critical for pulsation models, and for more flexibility in varying element abundance mixtures.  Commonly used equations of state are the analytical CEFF (Christensen-Dalsgaard \& Dappen 1992); the OPAL (Rogers et al. 1996) and MHD (Mihalas et al. 1988; Dappen et al. 1988) tables , and the SIREFF analytical treatment (see Guzik \& Swenson 1997).  There is also a sophisticated analytical EOS that includes excited states available from A. Irwin (see Cassisi et al. 2003) called FreeEOS available on-line. 


\subsection{Abundances and diffusive settling}

Stars are assumed to form with a homogeneous composition that is modified by nuclear reactions, element settling, radiative levitation, and mass loss during their evolution.  For normal stars a fixed mass fraction (Z) of elements heavier than hydrogen (X) and helium (Y)  is usually adopted, with the relative proportion of these elements scaled to the solar mixture.  Recently a major change to the solar abundance mixture has been proposed.   Asplund et al.'s (2005) analysis of the sun's spectrum using 3D dynamical model atmospheres, updated atomic physics, and non-LTE effects revised downward the solar Z, and in particular the fractions of the abundant elements O, C, N, Ne relative to Fe.  The Asplund et al. (2009) solar photosphere Z/X is 0.0181, whereas the Grevesse \& Noels (1993) value is 0.0244.   These mixtures, as well as the overall Z adopted, have a significant effect on stellar structure and oscillation properties.  Solar models calibrated with the newest abundance determinations unfortunately do not agree as well with solar oscillation constraints  (see Asplund et al. 2009; Basu \& Antia 2008; Guzik \& Mussack 2010), and so stellar modelers have been reluctant to adopt these new abundances in spite of the improved physics.

Diffusive settling processes (thermal diffusion, gravitational setting, and concentration diffusion), as well as radiative levitation (particularly of Fe) are also important in the evolution and pulsation modeling of some stars.  Descriptions and implementations of these processes in stellar models can be found in Thoul et al. (1994), Turcotte (1998a,b), Cox et al. (1989), Burgers (1969), Paquette et al. (1986a,b), and Iben \& MacDonald (1985).  For the sun, $\sim$5\% of the He and $\sim$5-10\% of the heavier elements settle out of the convection zone during its 4.5 Gyr lifetime, and produce a composition gradient near the convection zone base that significantly affects oscillation frequencies of calibrated solar models (see., e.g., Guzik et al. 2005).  Radiative levitation produces abundance anomalies in  main sequence A and F stars (Turcotte 1998b), and concentration of Fe due to levitation + settling is proposed as the mechanism for opacity-induced pulsation driving in the newly discovered classes of helium-burning subdwarf O and B variable stars (see, e.g., Randall et al. 2009; Rodr\'iguez-L\'opez et al. 2010; Hu et al. 2009, 2010).  Diffusion affects stellar age determinations and the location of the main sequence turnoff used for cluster dating, and is included in model grids for the lower main sequence (Dotter et al. 2008; Gai et al. 2009).   Gravitational settling stratifies white dwarf star envelopes into H, He, and C/O layers, and removes all of the metals from visibility on short timescales.  Composition gradient buildups can also cause instabilities and mixing that are now taken into account in some stellar models (e.g., Theado \& Vauclair 2009).

\subsection{Nuclear reaction rates}

Nuclear reaction rates for normal evolution stages have been taken from Caughlan \& Fowler (1988), but recent updates by Adelberger et al. (1998) and Angulo et al. (1999, NACRE) are now in general use.  Since then, there have been proposed updates to important reaction rates.  For example, using the exact $S$ factors for $^3$He+$^3$He, $^3$He+$^4$He, and $^7$Be+p of Kassim et al. (2010) decreases solar neutrino fluxes over those obtained using the NACRE rates by 6\% and 16\% for $^7$Be and $^8$B neutrinos, respectively.  The $^{14}$N+p $S$ factor was decreased from the NACRE rate by nearly  a factor of two (Formicola 2004), which reduced CNO-cycle solar neutrino output by a factor of two, and the ages of globular clusters by about 1 Gyr.   A new  theoretical He triple-alpha reaction proposed by  Ogata et al. (2009) was assessed by Dotter \& Paxton (2009); it produces a very early Helium flash and causes stars to bypass the ascent of the first giant branch!  Despite the maturity of this field, reaction rates seem far from settled.

In addition, the treatment of Coulomb screening significantly modifies reaction rates.  Most modelers have adopted the Salpeter (1954) approximation for static Coulomb screening.   However, it is the most energetic ions that engage in nuclear reactions, and recently dynamical screening corrections have been examined (see paper by Mussack, these proceedings and Shaviv \& Shaviv 2001).   Ions engaging in nuclear reactions are moving faster, so may not be accompanied by their full screening cloud.  Molecular dynamics calculations show that, for the p+p reaction in the solar core,  the reaction rates are closer to those obtained if no screening correction is applied.  Further investigation of this topic is needed, and clearly has implications for all of stellar evolution.

\section{Two and three-dimensional stellar interior models}

While most stellar evolution calculations have been one-dimensional, several groups have developed 2D and 3D tools to study aspects of stellar structure and evolution.

The ROTORC code of Deupree (1998, 2001, 2004) is a 2 1/2 D implict code to study core convection rotation and nuclear burning in massive stars with convective cores or shells. The code can be run in an evolution mode with large timesteps and the dynamics included in a parameterized way, and discontinued at any point to be run in a hydrodynamic mode to examine the extent of flows/mixing, and modify the parameterization for continued evolution.   Figure 2 shows a dynamical simulation of an  8.75 M$_{\odot}$ rotating main sequence model (Deupree 1998).  The arrows show instantaneous flow velocities.   Note that some material flows beyond the beyond the edge of the formal convective boundary.  For the rotating model only, the flow patterns align along cylinders in the convective core; without rotation, they are randomized more spherically.  Flows also show latitude dependence in the presence of rotation.   Figure 3 (Deupree 2004) shows tracer particle paths for a 20 M$_{\odot}$ shell hydrogen-burning rapidly rotating model.   Near the equator,  the mixing is more extended in radius and the timescales for circulation are faster than near the poles.
 \begin{figure}
 \includegraphics[width=\columnwidth]{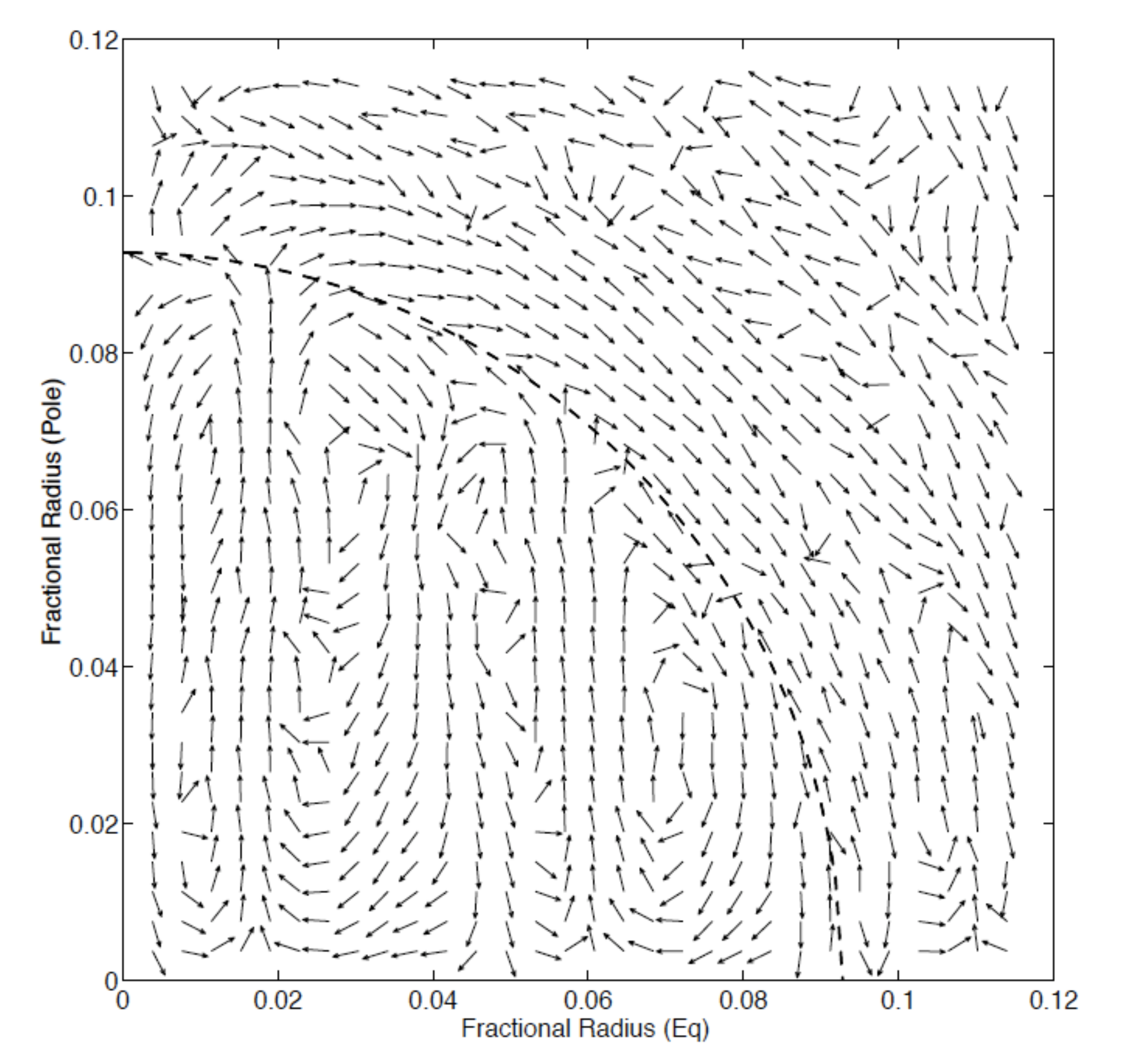}
 \caption{Instantaneous flow pattern for 8.75 M$_{\odot}$ rotating core hydrogen-burning model (Depuree 1998).} 
 \label{fig:2}
\end{figure}

 \begin{figure}
 \includegraphics[width=\columnwidth]{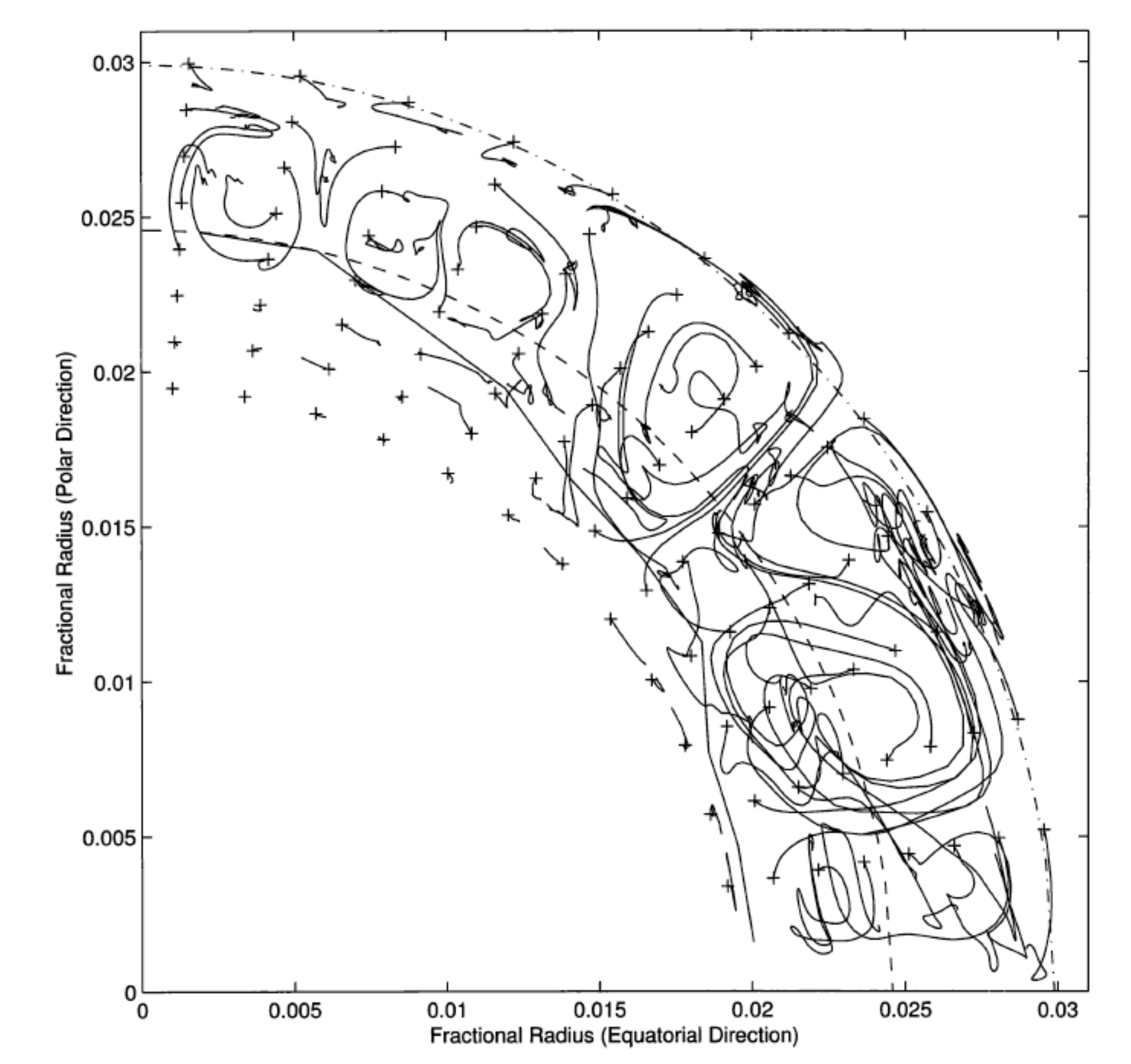}
 \caption{Instantaneous flow pattern for 20 M$_{\odot}$ rapidly rotating shell hydrogen burning model (Deupree 2004).} 
 \label{fig:3}
\end{figure}

Arnett et al. (2009, 2010) and Meakin \& Arnett (2007) have been modeling turbulent convection in stellar interiors with nuclear burning in 2D and 3D.  They propose a different picture of overshoot based on their results,  `turbulent entrainment', as opposed to ballistic penetration of turbulent eddies.   They studied H core burning and O shell burning.  They find significant difference in results and velocities scale and extent of mixing in 3D vs. 2D (Fig. 4), leading one to be cautions about the realism of parameterizing models based on 2D results.

Dearborn and collaborators have developed the Djehuty code at Lawrence Livermore National Laboratory for 3D stellar models. Djehuty is a 3D explicit code with nuclear burning, and can follow convection dynamics.  As a by-product of their work studying the Helium flash (see below), the Djehuty simulations revealed a layer unstable to mixing outside the hydrogen-burning shell at a small molecular weight inversion produced by $^3$He($^3$He,2$p$)$^4$He reaction (see Fig. 5) .  This mixing and burning eliminates the overproduction of $^3$He during the main sequence phase, and also explains the dredge-up of CNO-cycle products during the red giant branch ascent (Eggleton et al. 2008; Dearborn et al. 2006). 

At least three groups, Moc\'ak et al. (2008, 2009);  Dearborn et al. (2006); Deupree (1996) and (surprisingly early for multidimensional stellar modeling) Deupree (1984a, b) and Deupree \& Cole (1983); studied in 2D or 3D the important process of the core He flash, the ignition of core helium under degenerate conditions after the star has ascended the red giant branch.    They find that convection occurring with the onset of burning prevents explosion or large-scale alteration of the stellar structure during the flash, and keeps the star in hydrostatic equilibrium.  The details of the He flash are important for understanding the origin of hydrogen-deficient stars, mass loss on the asymptotic giant branch and during the He flash itself, horizontal branch morphology and luminosity.  Moc\'ak et al. (2008) showed by comparing 2D and 3D simulations that there is a significant difference in velocities and the extent of mixing in 3D compared to 2D, with smaller scale structures and lower velocities occurring in 3D.

Li et al. (2006, 2009) and Ventura (2009) have generalized the Yale Rotating Evolution code (YREC) to 2D for the study of stellar envelopes.  They include arbitrary magnetic fields, turbulence and rotation.  Their second paper modifies the treatment of turbulence and rotation to study short timescale phenomena ($< $1 year).  Their aim is to model effects of dynamo-type fields and rotation to high enough precision for comparisons with solar oscillation data.  Their work shows promise for producing and sustaining the differential rotation in the solar envelope inferred from seismic data.

Brun \& Toomre (2002), Brun \& Zahn (2006), Palacios et al. (2006), Mathis et al. (2006), and Brun \& Palacios (2009) have produced 3D models of envelope convection zones with their anelastic spherical harmonic code (ASH).  This code is semi-implicit, includes rotation, turbulence and magnetic fields.  They have applied it to the solar convection zone and red giants.  They find that they can sustain solar differential rotation, but do not produce the solar dynamo.  Their semi-implicit approach allows mean flows and meridional circulation occurring on longer timescales to be followed.

 \begin{figure}
 \includegraphics[width=\columnwidth]{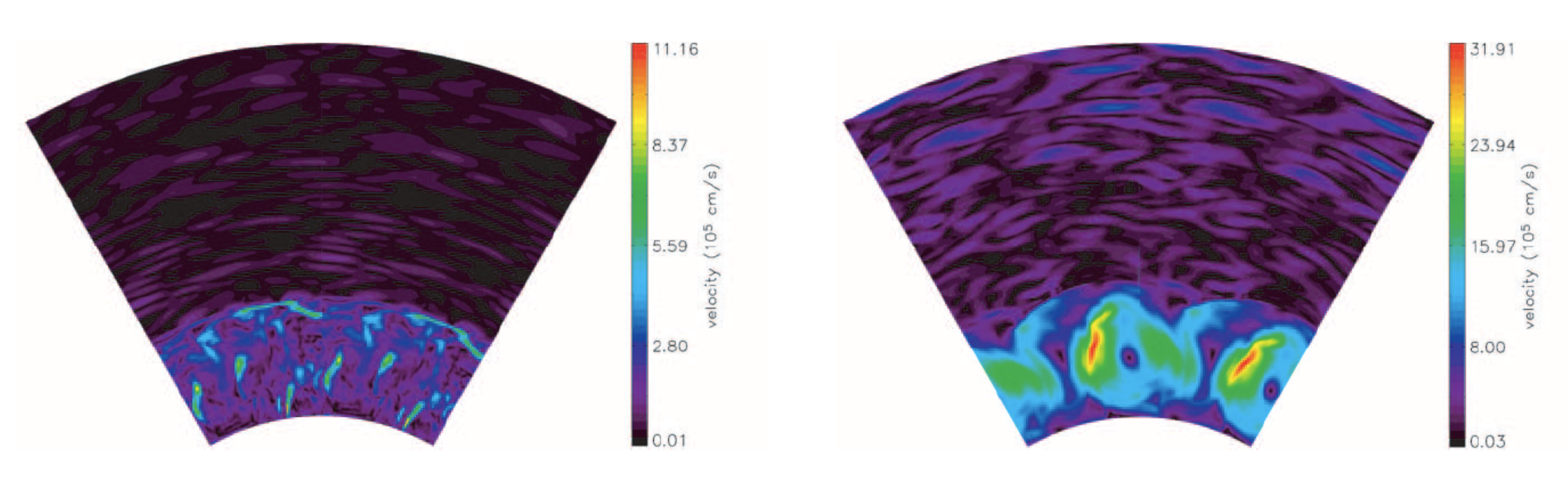}
 \caption{Velocity field in convective core of H burning stellar model in 2D (left) and 3D (right), showing larger scale eddies and faster velocities in 2D compared to 3D (color velocity scale increases x3 for 2D simulation)  (From Meakin \& Arnett 2007).} 
\label{fig:4}
\end{figure}

 \begin{figure}
 \includegraphics[width=\columnwidth]{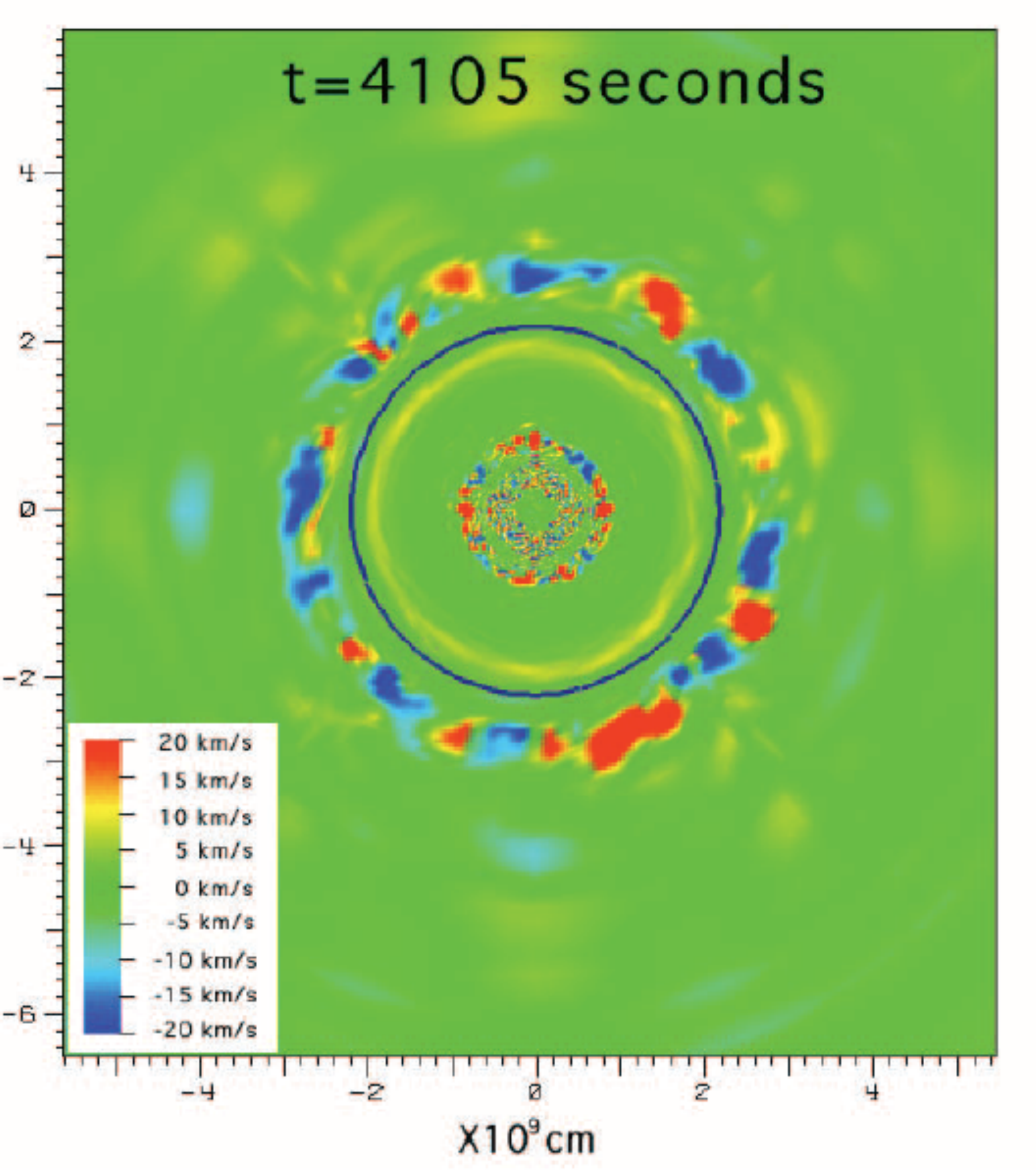}
 \caption{3D Djehuty simulation of He flash models shows  instability due to a molecular weight inversion (largest red and blue patches) produced by $^3$He+$^3$He reactions outside the hydrogen-burning shell (solid blue line). Color-coding is by radial velocity. (From Dearborn et al. 2006.)} 
 \label{fig:5}
\end{figure}

\section{Insights from stellar oscillation data and unsolved problems}

New asteroseismic data is arriving from space and ground-based networks.  Some of the major ones are the NASA Kepler spacecraft monitoring 100,000 stars (Gilliland et al. 2010);
CoRoT (Convection, Rotation and Planetary Transits, monitoring 11,400 stars; Baglin et al. 2009); the Canadian MOST spacecraft focusing on bright stars and sunlike stars (Matthews 2008); ; OGLE- III (1 million variable stars in the LMC; Soszy\'nski 2009); and  ASAS (All-Sky Automated Survey, that discovered 300 $\beta$ Cep stars (Pigulski \& Pojma\'nski 2009).

Instability domains and asteroseismology of $\beta$ Cep and SPB stars and their hybrids show sensitivity to abundances, abundance mixtures, and to OP vs. OPAL opacities (Zdravkov \& Pamyatnykh 2008a,b; Daszy\'nska-Dasczkiewicz \& Walczak 2009, 2010; Lenz et al. 2008). For the $\beta$ Cep/SPB hybrid $\gamma$ Peg, Zdravkov \& Pamyatnykh (2009) note that a 20-50\% opacity increase in OP opacities in the Fe bump regions at 200,000 K and 2 million K would reconcile observations with models.

Including effects of turbulence in outer layers is necessary to match 12 out of 13 radial mode frequencies observed by MOST for $\eta$ Boo (Straka et al. 2006).   SPB and $\beta$ Cep g- and p- modes may be able to distinguish between overshooting and rotational mixing at the convective core boundary (Miglio et al. 2008).  Signatures of overshooting and semiconvection may leave detectable asteroseismic signatures (Noels et al. 2010).  Asteroseismology may also be able to constrain internal differential rotation (Su\'arez et al. 2010)

We close by pointing out two more unsolved problems, in addition to the solar abundance problem noted above:  We do not understand why nearly all of the hundreds of $\gamma$ Dor and $\delta$ Sct stars observed by Kepler appear to be hybrids, pulsating with both long $\gamma$ Dor and short $\delta$ Sct periods, while theory predicts that there should be two distinct classes and that hybrids should exist only in a small overlapping region of their instability strips in the HR diagram (Grigahc\'ene et al. 2010).  We also do not know the cause of outbursts accompanied by substantial mass loss in luminous blue variable stars, including the giant eruptions seen in $\eta$ Car and P Cyg (see, e.g., Gonz\'alez et al. 2010; Vink 2009; Smith 2008)

\section{Summary and conclusions}

2D and 3D models are now becoming sophisticated enough to study (and reveal neglected) dynamical processes in stars.  In many cases 3D models generate significantly different results than 2D models.  The 2D and 3D models are being used to generate refinements in parameterized 1D models for longer-timescale evolution calculations.  Updates continue to be made in input and physics for stellar models:  abundances, opacities, EOS, nuclear reaction rates, electron screening, diffusive settling.  Observations from the ground and space are generating profuse data on stellar oscillations that are being used to test the physics and input for stellar interior models.  Many outstanding problems remain, and new ones  revealed by new data on stellar oscillations are emerging.  The research of understanding and simulating the normal evolution and oscillations of single stars is far from complete.

%

%
%

%

%

%
\acknowledgments
J.G. appreciates the assistance of K. Mussack, D. Dearborn, D. Arnett, R. Deupree, A.N. Cox, P. Bradley, and S. Turck-Chieze in preparation of this paper, as well as the invitation from W. Dappen, T. Plewa, and  R. Mancini.


%
\bibliographystyle{Spr-mp-nameyear-cnd.bst}  
\bibliography{}          




{}

\end{document}